\documentclass[footinbib,pra,twocolumn,superscriptaddress]{revtex4-1}

\usepackage{pbox}
\usepackage{graphicx}
\usepackage{amsmath}
\usepackage{braket}
\usepackage{float}

\usepackage{textcomp}
\usepackage{tabularx,ragged2e,booktabs,caption}

\usepackage{amsthm}

\usepackage{subfig}
\usepackage{url}

\begin{document}
	\title{Decomposed description of Ramsey spectra under atomic interactions}
	\author{Ryotatsu Yanagimoto}
	\email[Email: ]{ryotatsu@stanford.edu}
	\altaffiliation[now ]{Edward L. Ginzton Laboratory, Stanford University, Stanford, California 94305, USA}
	\affiliation{Quantum Metrology Laboratory, RIKEN, Wako-shi, Saitama 351-0198, Japan}
	\affiliation{Department of Applied Physics, Graduate School of Engineering, The University of Tokyo, Bunkyo-ku, Tokyo 113-8656, Japan}
	\author{Nils Nemitz}
	\altaffiliation[Now ]{Space-Time Standards Laboratory, NICT, Koganei, Tokyo 184-0015, Japan}
	\affiliation{RIKEN Center for Advanced Photonics, Wako-shi, Saitama 351-0198, Japan}
	\author{Filippo Bregolin}
	\affiliation{Quantum Metrology Laboratory, RIKEN, Wako-shi, Saitama 351-0198, Japan}
	\affiliation{Physical Metrology Division, Istituto Nazionale di Ricerca Metrologica (INRIM), Strada delle Cacce 91, 10135 Torino, Italy}
	\affiliation{Dipartimento di Elettronica e Telecomunicazioni, Politecnico di Torino, Corso duca degli Abruzzi 24, 10129 Torino, Italy}
	\author{Hidetoshi Katori}
	\affiliation{Quantum Metrology Laboratory, RIKEN, Wako-shi, Saitama 351-0198, Japan}
	\affiliation{Department of Applied Physics, Graduate School of Engineering, The University of Tokyo, Bunkyo-ku, Tokyo 113-8656, Japan}
	\date{\today}
\begin{abstract}
We introduce a description of Ramsey spectra under atomic interactions as a sum of decomposed components with differing dependence on interaction parameters. This description enables intuitive understanding of the loss of contrast and asymmetry of Ramsey spectra. We derive a quantitative relationship between the asymmetry and atomic interaction parameters, which enables their characterization without changing atom density. The model is confirmed through experiments with a Yb optical lattice clock.
\end{abstract}
	\maketitle

\section{Introduction}
\begin{figure}[h]
	\captionsetup{singlelinecheck = false, justification=raggedright, font=footnotesize, labelsep=space,position=top}

	\includegraphics[width=0.4\textwidth]{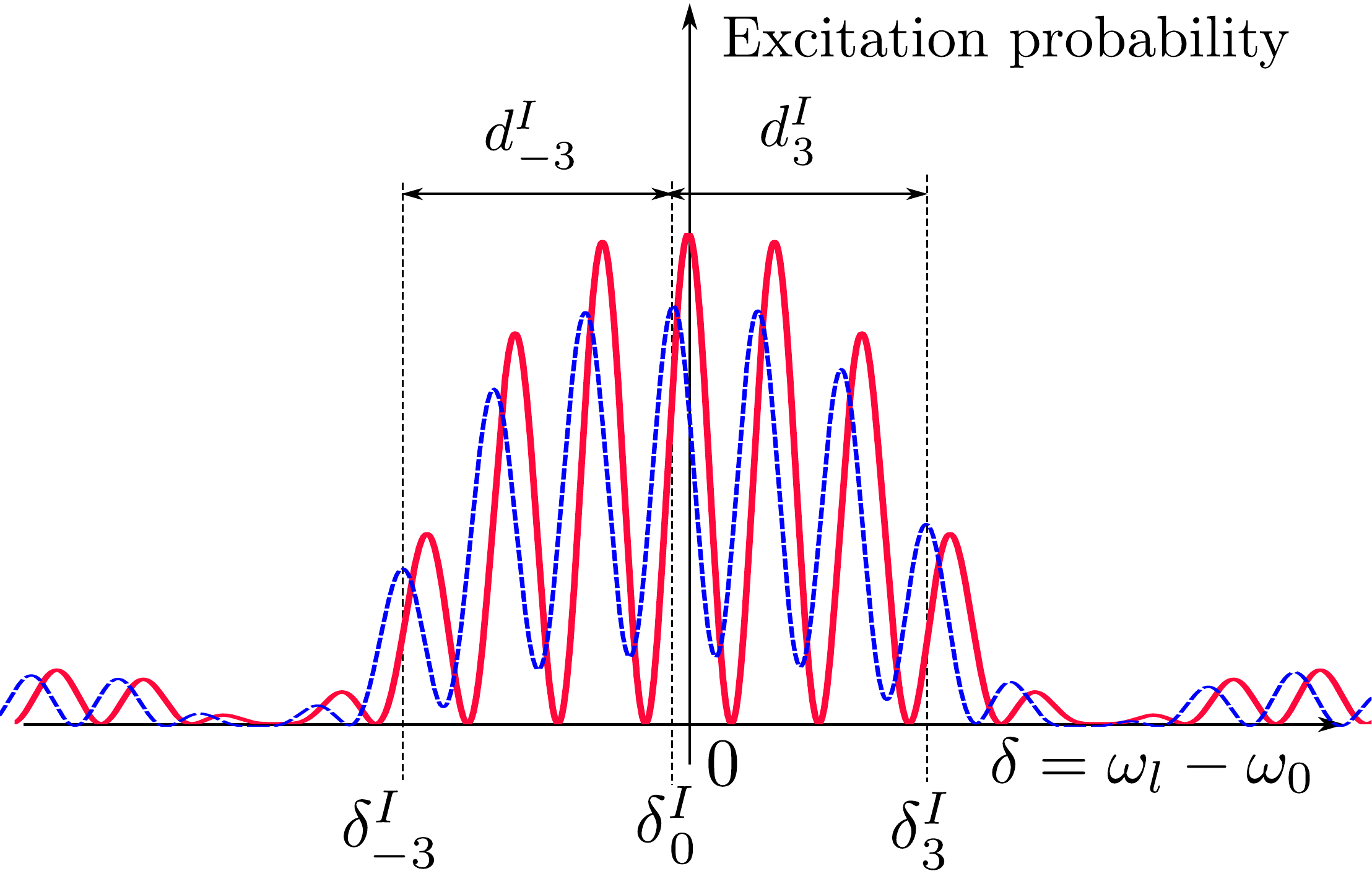}

	\caption{Ramsey spectrum with (blue dashed line) and without (red solid line) atomic interactions. Without interactions, the spectum is symmetric about $\delta=0$. Atomic interactions shift the $n^\text{th}$ peak to $\delta_n^\text{I}$. The separations from the center fringe $d_n^\text{I}$ and $d_{-n}^\text{I}$ then become unequal, resulting in asymmetry. As an example, $d_{3}^\text{I}$ and $d_{-3}^\text{I}$ are shown. Plotted spectrum is obtained by a sequence of two $\tau=16$ ms $\frac{\pi}{2}$-pulses separated by a $60$ ms dark interval.}
	\label{schematics}
	
\end{figure}

Ramsey spectroscopy is one of the standard techniques of precision measurements of atomic resonances \cite{ramsey1950}. It employs two excitation pulses, typically of equal length $\tau$, that are separated by a dark interval, where atoms are in a freely evolving quantum superposition state. The resulting excitation probability as a function of the frequency of the exciting field shows characteristic spectra as in Figure~\ref{schematics}. Compared to Rabi spectroscopy where atoms are exposed to a single, continuous pulse, Ramsey spectroscopy is capable of providing a reduction in linewidth by a factor of $1.7$ for a given interrogation time \cite{ramsey1950,torrey1941}. It is also possible to extend the Ramsey method to achieve better controls of atomic states. An example for this is the Hyper-Ramsey scheme \cite{yudin2010}, where the addition of a third pulse, along with careful control of amplitude and phase, can eliminate frequency shifts resulting from the excitation field itself \cite{sanner2017}.

In a situation where multiple atoms are interrogated simultaneously, their interactions are of significant importance for precision measurements. For example, the resulting frequency shifts in cesium fountain clocks need to be either measured continuously and with great accuracy \cite{santos2002}, or the conditions governing the interactions have to be precisely controlled to minimize their effect \cite{szymaniec2007}.

When $p$-wave atomic interactions (described by anti-symmetric wavefunctions) are dominant, their dependence on the excitation probability can be used to control the collisional shifts \cite{lemke2011,lee2016}. This control of collisional shift is experimentally feasible in certain systems of ultra-cold atoms, such as neutral ${}^{171}\text{Yb}$, where the contribution caused by $p$-wave atomic interactions is sufficiently larger than that of $s$-wave atomic interactions (described by symmetric wavefunctions). Our work focuses on situations where this assumption of $p$-wave dominance is valid.

The shift of the central peak is usually the most relevant concern for precision measurements \cite{hazlett2013,maineult2012}, while atomic interactions also alter the shape of the entire spectrum, manifesting as loss of contrast and asymmetry as discussed in Ref.~\cite{Band2006}. Here, we present a model that decomposes the Ramsey spectrum into the sum of two components distinguished by their dependence on the interaction parameters. This intuitively describes the loss of contrast due to atomic interactions through the simultaneous presence of both components. We also derive a formula which quantitatively relates the summed interaction parameter $W$, which will be discussed in detail later, to the asymmetry of the Ramsey spectra. A measurement method based on this allows measuring the strength of atomic interactions without changing the atom density. The new model is experimentally confirmed using our ${}^{171}\text{Yb}$ optical lattice clock.  

\section{Theoretical model}
We consider spin-polarized fermionic atoms trapped inside a 1D optical lattice oriented along the $z$ axis. The lattice is created by the standing wave of a laser at a magic frequency minimizing the ac Stark shift of the clock transition \cite{katori2003}. The tight confinement along the lattice axis allows interrogation of atoms in the Lamb-Dicke regime for the co-propagating clock laser.

As described in Ref.~\cite{lemke2011}, we first consider a simple case where two fermions $1,2$ are trapped inside a lattice site. We assume that these fermions are spin polarized along the direction of a homogenous external magnetic field $B$. We denote the vibrational quantum number of atom $i = 1,2$ along the $j = x,y,z$ direction as $n_{ij}$.

\begin{figure}[h]
	\captionsetup{singlelinecheck = false, justification=raggedright, font=footnotesize, labelsep=space,position=top}
		\includegraphics[width=0.4\textwidth]{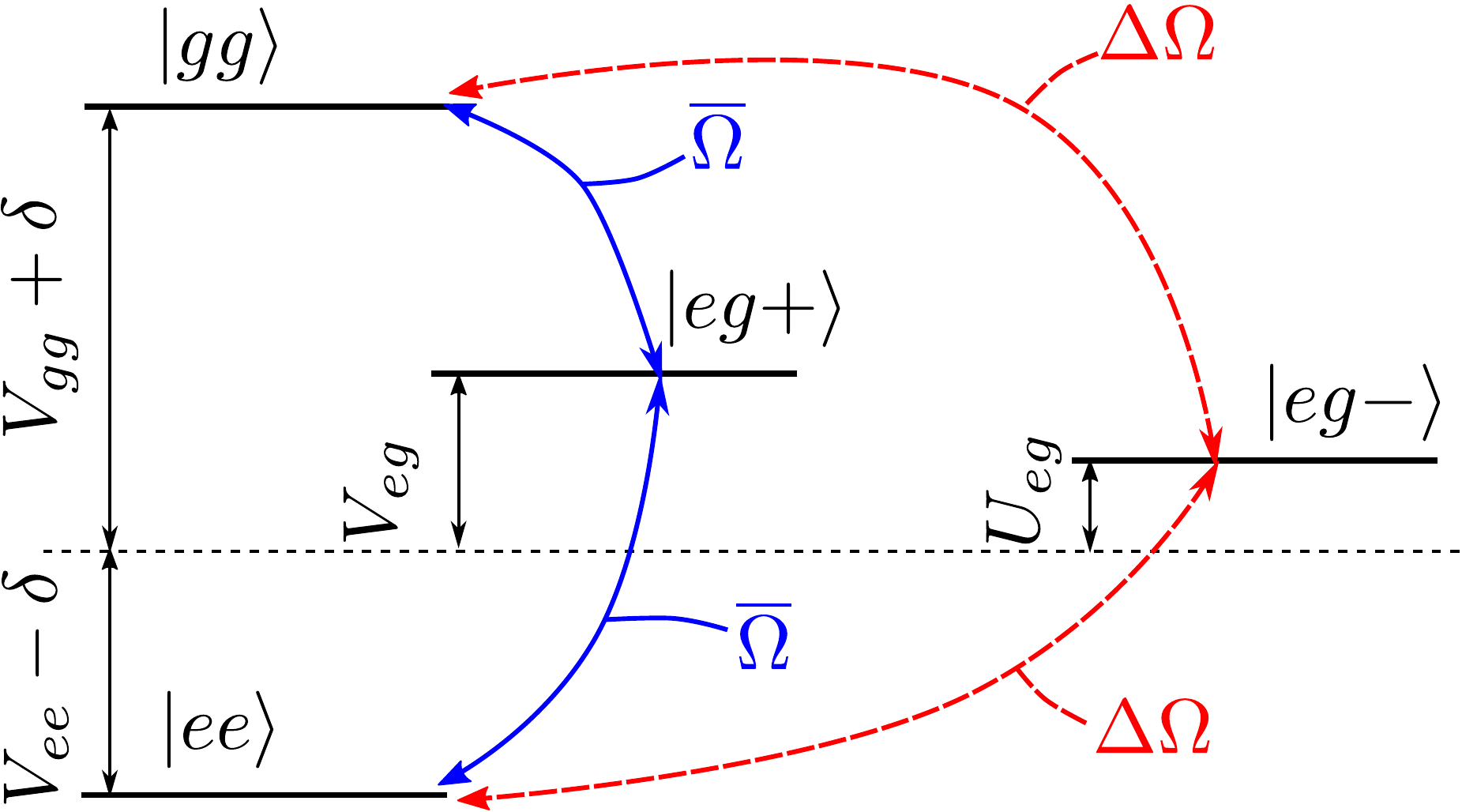}

	\caption{Energy shifts of two-atom states due to the atomic interactions whose strength is denoted by interaction parameters $V^{\alpha\beta}$ and $U^{\alpha\beta}$. Straight arrows show energy differences in terms of detuning $\delta$, and curved arrows indicate coupling strengths expressed as Rabi frequencies.}
	\label{models}
\end{figure}

The state of the two atoms can be written in a four-state basis composed of (in order) three triplet states $\ket{gg}$, $\ket{ee}$, and $\ket{eg+}=\left(\ket{eg}+\ket{ge}\right)/\sqrt{2}$ and a singlet state $\ket{eg-}=\left(\ket{eg}-\ket{ge}\right)/\sqrt{2}$, depending on whether each atom is in the electronic ground state $\ket{g}$ or in the excited state $\ket{e}$. Since the overall wavefunction of spin-polarized fermions has to be anti-symmetric about particle exchange, atom pairs in the triplet and the singlet electric states have anti-symmetric and symmetric spatial wavefunctions respectively. We will therefore refer to their lowest order interactions as ”$p$-wave” and ”$s$-wave” interactions, since higher order interactions are suppressed due to the low temperature of the atoms. Respectively, we denote the corresponding energy shifts as interaction parameters $V^{\alpha\beta}$ and $U^{\alpha\beta}$ for atomic states $\alpha=g,e$ and $\beta=g,e$~\cite{lemke2011,rey2014} as shown in Figure~\ref{models}. In the presence of an electromagnetic field of detuning $\delta=\omega_l-\omega_0$ from the atomic resonance $\omega_0$, the two-body Hamiltonian in the four-state basis becomes \cite{lemke2011}
	\begin{equation}\label{hamiltonian}
\hat{H}=
\left(
\begin{array}{cccc}
\delta+V^{gg}&0&\overline{\Omega}/\sqrt{2}&\Delta\Omega/\sqrt{2}\\
0&-\delta+V^{ee}&\overline{\Omega}/\sqrt{2}&-\Delta\Omega/\sqrt{2}\\
\overline{\Omega}/\sqrt{2}&\overline{\Omega}/\sqrt{2}&V^{eg}&0\\
\Delta\Omega/\sqrt{2}&-\Delta\Omega/\sqrt{2}&0&U^{eg}\\
\end{array}
\right).
\end{equation}
where $\overline{\Omega} = (\Omega_1 + \Omega_2)/2$ and $\Delta\Omega = (\Omega_1- \Omega_2)/2$ are the mean and deviation of the Rabi frequencies experienced by the two atoms. $\Delta\Omega/\overline{\Omega}$ corresponds to the inhomogeneity of the Rabi frequency of the atoms. As the contribution from $s$-wave interactions is on the order of $\mathcal{O}\left(\frac{\Delta\Omega^2}{\overline{\Omega}^2}\right)$, the contribution becomes negligible when the Rabi frequency of the atoms is homogeneous. However, it is generally difficult to completely eliminate $\Delta\Omega$, since any misalignment of the clock laser from the axis of strong confinement causes the Rabi frequency to depend on the radial vibrational modes $n_x$ and $n_y$ \cite{takamoto2009}.

\begin{figure}[h]
	\captionsetup{singlelinecheck = false, justification=raggedright, font=footnotesize, labelsep=space,position=top}

	\subfloat[]{
		\includegraphics[width=0.47\textwidth]{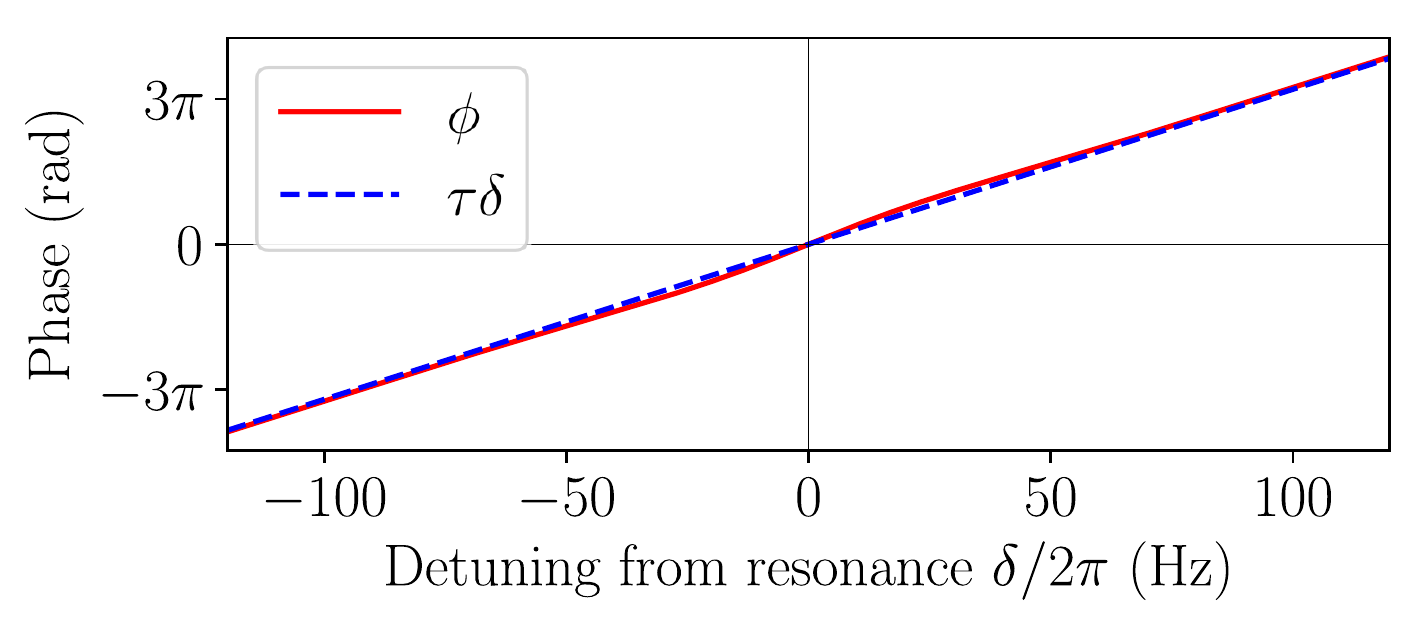}
		\label{phase}
	}\hfill
	\subfloat[]{
	\includegraphics[width=0.49\textwidth]{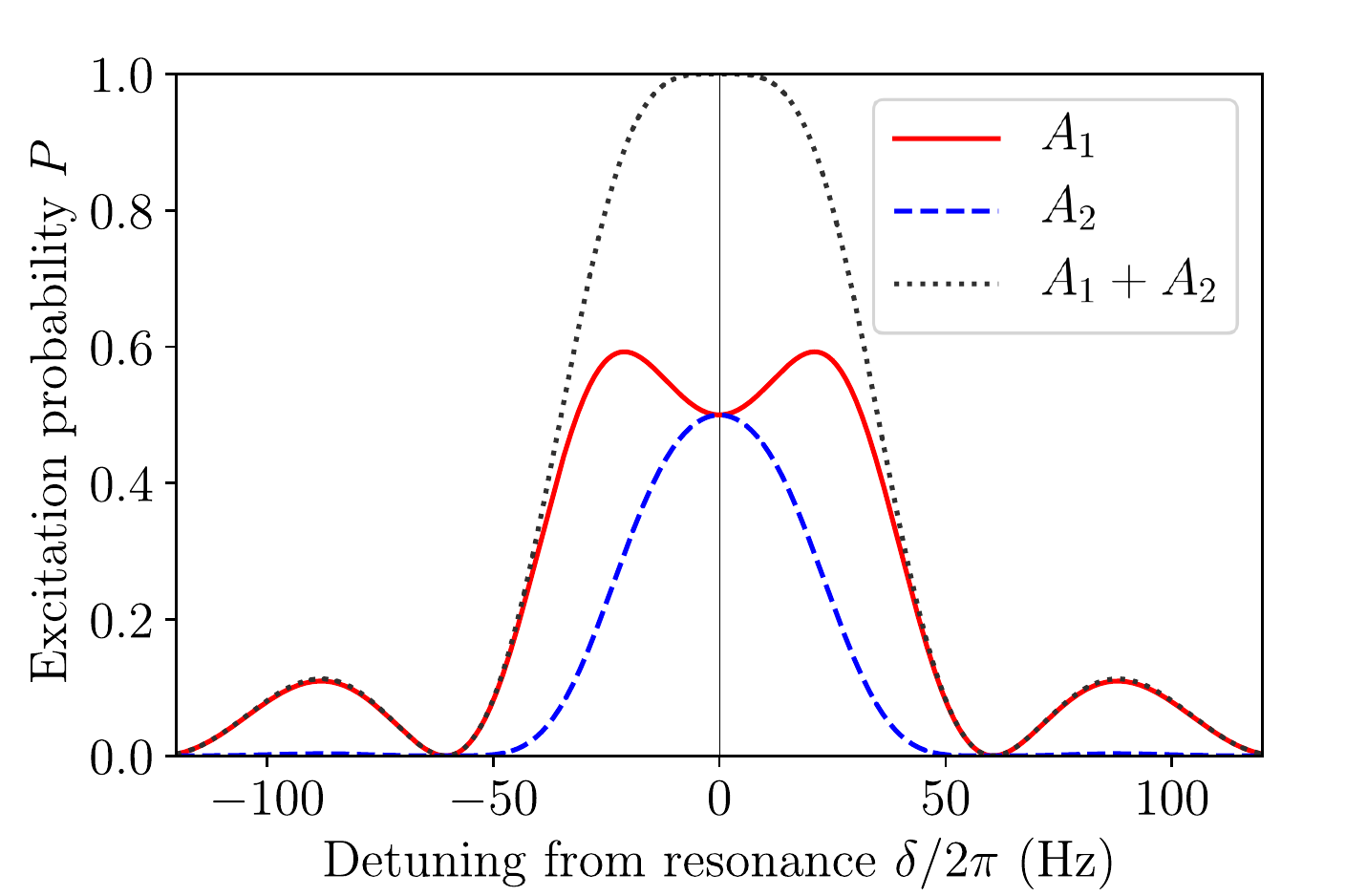}		
	\label{a1a2}
}
	\caption{(a) $\phi$ (red solid line) plotted together with the approximation $\delta\tau$ (blue dashed line) for $\tau=16$ ms $\frac{\pi}{2}$-pulse. (b) $A_1$ and $A_2$, plotted as red solid and blue dashed lines respectively for the same parameters as (a). As a reference, $A_1+A_2$, which corresponds to the envelope of the Ramsey spectra in the absence of atomic interactions, is also plotted as dotted gray lines.}
	\label{outline}
\end{figure}

The Rabi frequency is treated as constant during the excitation pulses, which are characterized by a pulse area $\overline{\Omega}\tau$. For atoms initially in the state $\ket{gg}$, we use the Hamiltonian in Eq.~(\ref{hamiltonian}) to calculate the excitation probability $P(\delta)$ after a Ramsey sequence consisting of two identical pulses separated by a dark time $T$, during which $\overline{\Omega}$ and $\Delta\Omega$ are zero. For small inhomogeneity $\Delta\Omega /\overline{\Omega}$, $P(\delta)$ can be decomposed into a sum of two oscillating components as
	\begin{equation}\label{spectrum}
	\begin{split}
	P(\delta)&=A_1(\delta)\cos^2\left[\frac{\left(\delta-V^{eg}+V^{gg}\right)T+\phi(\delta)}{2}\right]\\
	&+A_2(\delta)\cos^2\left[\frac{\left(\delta+V^{eg}-V^{ee}\right)T+\phi(\delta)}{2}\right]\\
	&+\mathcal{O}\left(\frac{\Delta\Omega^2}{\overline{\Omega}^2}\right)
	\end{split}
	\end{equation}
	with envelope functions
	\begin{equation}\label{a1}
	\begin{split}
	A_1(\delta)&=\frac{\overline{\Omega }^2 }{\left(\overline{\Omega }^2+\delta ^2\right)^3}\sin ^2\left(\frac{1}{2} \tau \sqrt{\overline{\Omega }^2+\delta ^2}\right) \times\\
	&\left\{\overline{\Omega }^2 \left[\cos \left(\tau \sqrt{\overline{\Omega }^2+\delta ^2}\right)+1\right]+2 \delta ^2\right\}^2,
	\end{split}
	\end{equation}
	\begin{equation}\label{a2}
	\begin{split}
	A_2(\delta)&=\frac{2 \overline{\Omega }^4 }{\left(\overline{\Omega }^2+\delta ^2\right)^3}\sin ^4\left(\frac{1}{2} \tau \sqrt{\overline{\Omega }^2+\delta ^2}\right) \times\\
	&\left\{\overline{\Omega }^2 \left[\cos \left(\tau \sqrt{\overline{\Omega }^2+\delta ^2}\right)+1\right]+2 \delta ^2\right\},
	\end{split}
	\end{equation}
	and the additional phase arising from the finite pulse length $\tau$
	\begin{equation}
	\begin{split}
	\phi(\delta)&=\arctan\left[\frac{2 \delta  \sqrt{\overline{\Omega }^2+\delta ^2} \sin \left(\tau \sqrt{\overline{\Omega }^2+\delta ^2}\right)}{\left(\overline{\Omega }^2+2 \delta ^2\right) \cos \left(\tau \sqrt{\overline{\Omega }^2+\delta ^2}\right)+\overline{\Omega}^2}\right],
	\end{split}
	\end{equation}
which is an odd function. As shown in Figure~\ref{outline}\subref{phase}, this additional phase can be well approximated as $\phi(\delta)\simeq\delta\tau$.   
In the limit of weak interactions ($V^{\alpha\beta}\rightarrow0$), the phases of the oscillating components given by the first and second line of Eq.~(\ref{spectrum}) are identical and $P(\delta)$ simplifies to the Ramsey spectrum without atomic interactions. Otherwise the $V^{eg}-V^{gg}$ and $-V^{eg}+V^{ee}$ terms introduce different phase shifts without affecting the envelope functions $A_1$ and $A_2$. Note that for $T\gg \tau$, the slow variation of $A_1$ and $A_2$ is negligible compared to the oscillation resulting from $\delta T$, and thus the oscillatory behavior of the spectrum is mostly explained by the cosinusoidal part of Eq.~(\ref{spectrum}). Physically, as can be seen from the phase of the cosinusoidal part, the $A_1$ term ($A_2$ term) is generated by the interference between the phase of $\ket{gg}$ and $\ket{eg+}$ states ($\ket{eg+}$ and $\ket{ee}$ states) during the dark time (see Figure~\ref{models}). Note that these envelope functions do not depend on the atomic interactions, or the length of the dark time $T$. For illustration, $A_1$ and $A_2$ are plotted in Figure~\ref{outline}\subref{a1a2} for a typical Ramsey sequence. While $A_1$ and $A_2$ have comparable contributions to the spectrum in the central part, $A_1$ starts to dominate over $A_2$ with increased detuning, which reflects the decrease in the population in the $\ket{ee}$ state. In general, the phase shifts of the two decomposed components are not equal ($V^{eg}-V^{gg}\ne -V^{eg}+V^{ee}$), resulting in a loss of contrast where both $A_1$ and $A_2$ exist. In a typical spectrum, this effect is most visible in the center region (Figure~\ref{schematics}).

We now discuss the frequency shift in the picture of the decomposed description. First, as shown in Figure~\ref{schematics}, we enumerate the peaks starting at $\delta=0$. We also denote the $n^\text{th}$ peak's original positions in the absence of atomic interactions as $\delta_{n}$, for which $\phi(\delta)\sim\delta\tau$ yields $\delta_{n}\sim\frac{2\pi n}{T+\tau}$. In the vicinity of $\delta_{n}$, $P$ can be expanded as a quadratic function as
\begin{equation}
\begin{split}
P(\delta)\simeq& -\alpha\left[ A_1\left(\delta_{n}\right)\left(\delta-\delta_{n}-V^{eg}+V^{gg}\right)^2\right.\\
 &\left.+A_2\left(\delta_{n}\right)\left(\delta-\delta_{n}+V^{eg}-V^{ee}\right)^2\right]+\text{const}
 \end{split}
 \end{equation}
using a positive constant $\alpha$. It can be rewritten as
 \begin{equation}
P(\delta)\simeq-\alpha(A_1+A_2)\left(\delta-\delta_{n}^\text{I}\right)^2+\text{const},
\end{equation}
where $\delta_{n}^\text{I}$represents the position for the $n^\text{th}$ peak in the presence of atomic interactions. As a consequence, the frequency shift can be written as
\begin{small}
\begin{equation}\label{shiftformula}
\begin{split}
\delta_{n}^\text{I}-\delta_n=\frac{A_1(\delta_{n})\left(V^{eg}-V^{gg}\right)+A_2(\delta_{n})\left(-V^{eg}+V^{ee}\right)}{A_1(\delta_{n})+A_2(\delta_{n})}.
\end{split}
\end{equation}
\end{small}
As can be seen in the equation, $\delta_{n}^\text{I}-\delta_{n}$ is an average of the shift of each oscillating component weighted by their amplitudes. After some calculations, $\delta_{n}^\text{I}-\delta_{n}$ becomes
{\small
\begin{equation}\label{shiftformuladetailed}
\delta_{n}^\text{I}-\delta_{n}=\frac{V^{ee}-V^{gg}}{2}-\frac{\delta_{n}^2+\overline{\Omega}^2\cos\left(\tau\sqrt{\delta_{n}^2+\overline{\Omega}^2}\right)}{2\left(\delta_{n}^2+\overline{\Omega}^2\right)}W
\end{equation}}with $W=-2V^{eg}+V^{gg}+V^{ee}$. For the center peak in particular, the shift becomes
\begin{equation}\label{pwave}
\delta_{0}^\text{I}=\frac{V^{ee}-V^{gg}}{2}+\left(p_1-\frac{1}{2}\right)W,
\end{equation}
where $p_1$ is the excitation probability after the first pulse. This result is identical to that of Ref \cite{lemke2011}. It shows that $W$ is experimentally accessible by measuring $\delta_{0}^\text{I}$ for various values of $p_1$, and we shall refer to this as the ”$p_1$-based” measurement method.

In the following, we quantitatively formalize the asymmetry of the Ramsey spectrum using Eq.~(\ref{shiftformula}). When accounting for atomic interactions, the frequency separation $d_{n}^\text{I}$ from the $0^\text{th}$ peak is
	\begin{equation}\label{shiftplus}
	d_{n}^\text{I}=\left\lvert\delta_{n}^\text{I}-\delta_{0}^\text{I}\right\rvert.
	\end{equation}
As a measurable quantity for the asymmetry of the Ramsey spectrum for the $\pm n^\text{th}$ peaks, we define
	\begin{equation}\label{asymmetryformula}
	a_{n}=\frac{d_{n}^\text{I}-d_{-n}^\text{I}}{2}.
	\end{equation}
	By substituting Eq.~(\ref{shiftplus}) and using the result of Eq.~(\ref{shiftformuladetailed}) together with the anti-symmetry of the peak positions $\delta_{n}=-\delta_{-n}$, the asymmetry $a_{n}$ becomes
	\begin{equation}\label{asymmetryequation}
	\begin{split}
	a_{n}&=\frac{\delta_{n}^\text{I}+\delta_{-n}^\text{I}}{2}-\delta_{0}^\text{I}=C_{n}W,
	\end{split}
	\end{equation}
	where we have defined
	\begin{equation}
	C_{n}=\frac{1}{2}\left[\cos\left(\overline{\Omega}\tau\right)-\frac{\overline{\Omega}^2\cos\left(\tau\sqrt{\overline{\Omega}^2+\delta_{n}^2}\right)+\delta_{n}^2}{\overline{\Omega}^2+\delta_{n}^2}\right].
	\end{equation}
Since $C_{n}$ depends only on the known experimental quantities $\overline{\Omega }$ and $\delta_{n}$, it is possible to directly relate the measured asymmetry $a_{n}$ to $W$ using Eq.~(\ref{asymmetryequation}), and we shall refer to this method as the ”asymmetry-based” measurement method of $W$.\\
\section{Experimental confirmation}
\begin{figure}[h]
		\captionsetup{singlelinecheck = false, justification=raggedright, font=footnotesize, labelsep=space}
		\includegraphics[width=0.45\textwidth]{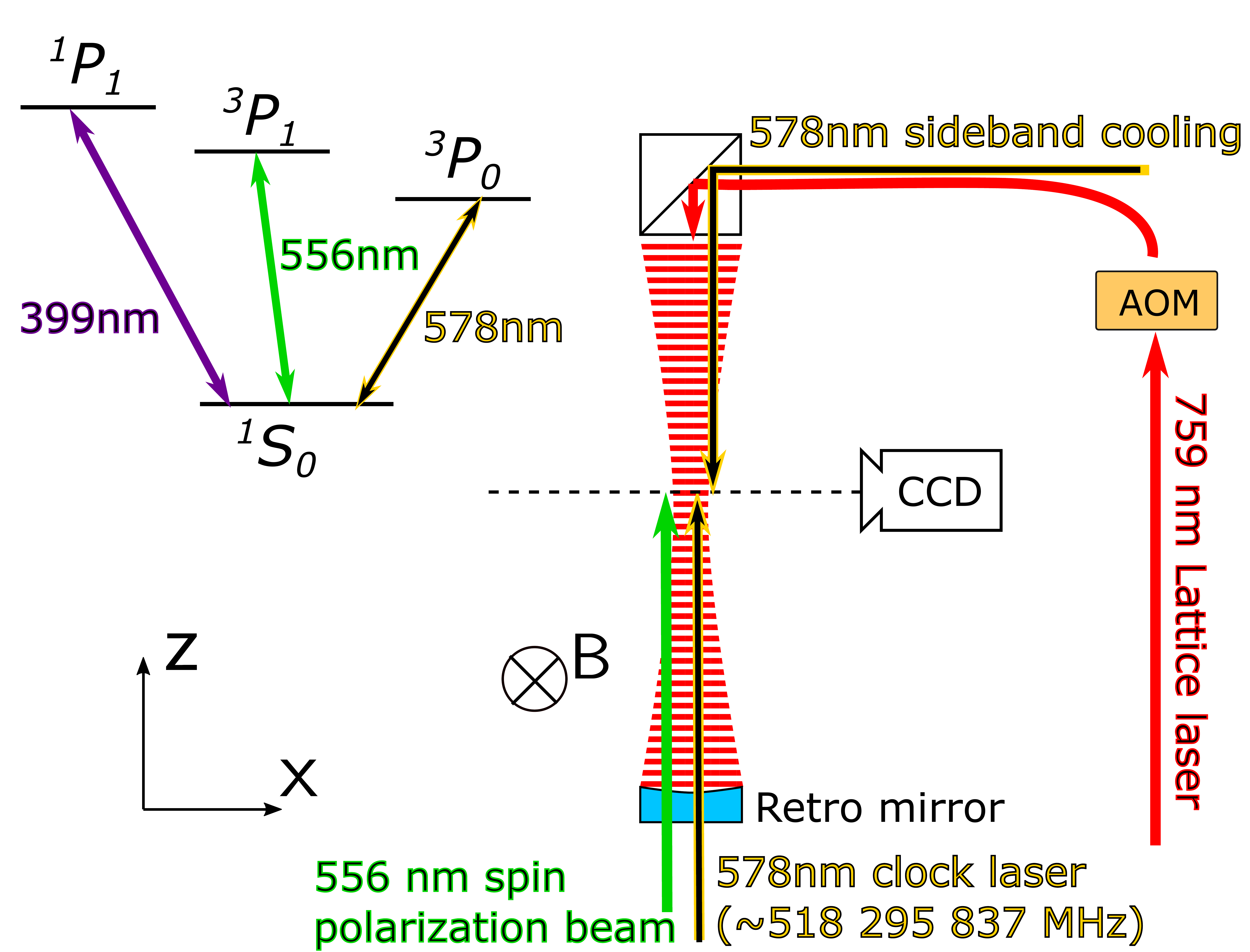}
		\caption{Overview of the experimental setup of our ${}^{171}\text{Yb}$ optical lattice clock. The upper left inset shows the relevant electronic states of the ${}^{171}\text{Yb}$ atom.}
		\label{chamber}	
	\end{figure}

To test the theoretical model, measurements were performed with the Yb optical lattice clock \cite{nemitz2016}. Figure~\ref{chamber} gives an overview of the experimental setup. Atoms are cooled down through two stages of magneto-optical trap and trapped in the magic wavelength optical lattice. This is created by a retro-reflected beam with a radius of $w\simeq 43$ \text{$\mu$}m  at the trap position.  Its intensity is actively stabilized using an acousto-optic modulator (AOM). 
The axial motional state $n_z$ is sideband cooled via the red sideband transition ${}^1S_0(n_z) \rightarrow {}^3P_0(n_z-1)$. The supression of the red sideband after the cooling sequence indicates that more than $95$\% of the atoms populate the axial vibrational ground state $n_z = 0$. Atoms are spin polarized in the $m_F=1/2$ or $-1/2$ state by optical pumping on the ${}^1S_0-{}^3P_1$ transition, reducing the population in the undesired spin state to less than $1$\%. The excitation probability after the clock laser pulses is determined by measuring the fluorescence on the ${}^1S_0-{}^1P_1$ transition using a CCD camera.

\begin{figure}[h]		\captionsetup{singlelinecheck = false, justification=raggedright, font=footnotesize, labelsep=space}	
	\includegraphics[width=0.5\textwidth]{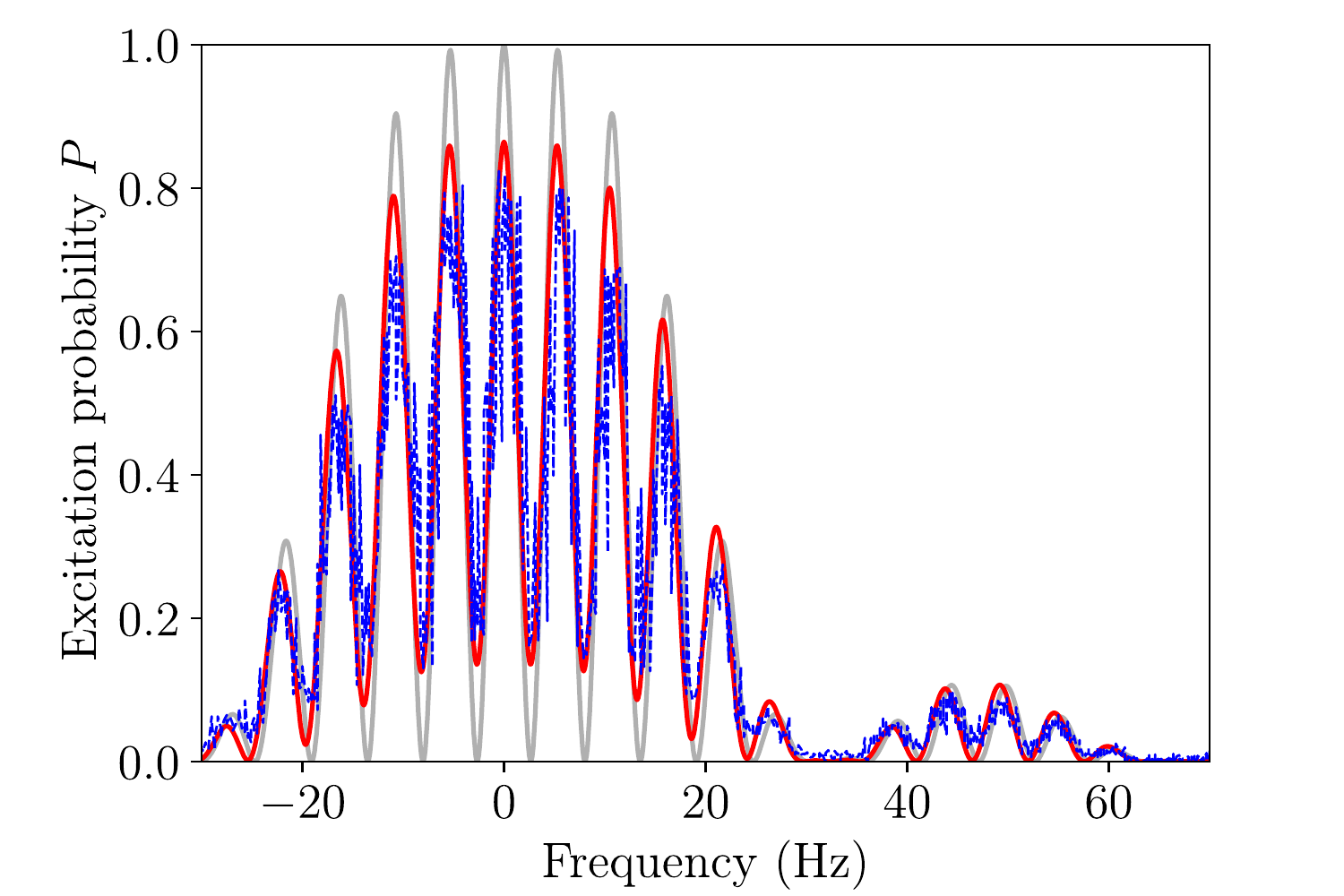}
	\caption{Experimental Ramsey spectrum (blue dashed line) for $\tau=30$ ms $\frac{\pi}{2}$ pulses and $T=150$ ms with an atom number of $\sim1500$, which corresponds to an average number of $1.5\sim3$ atoms per lattice site. To increase atomic interactions, the trap depth is set to a large value of $650E_\text{r}$. Gray line indicates the theoretical spectrum without atomic interactions. Red line includes interactions with $\widetilde{W}=2\pi\times1.6$ Hz obtained from experimental data by the asymmetry-based measurement.}
	\label{entirespectrum}
\end{figure}

Figure~\ref{entirespectrum} shows a Ramsey spectrum taken with a particularly strong confinement corresponding to a trap depth of $650E_\text{r}$, where $E_\text{r}=\frac{h^2}{2m\lambda^2}$ is the lattice photon recoil energy for an atom mass of $m$ and lattice wavelength $\lambda=759$ nm. The comparison to a theoretical spectrum calculated based on Eq.~(\ref{spectrum}), clearly shows that the model successfully reproduces the loss of contrast that is more pronounced in the center of the spectrum than in the side lobes.

\begin{figure}[h]
	\captionsetup{singlelinecheck = false, justification=raggedright, font=footnotesize, labelsep=space,position=top}
	\subfloat[]{
		\includegraphics[width=0.405\textwidth]{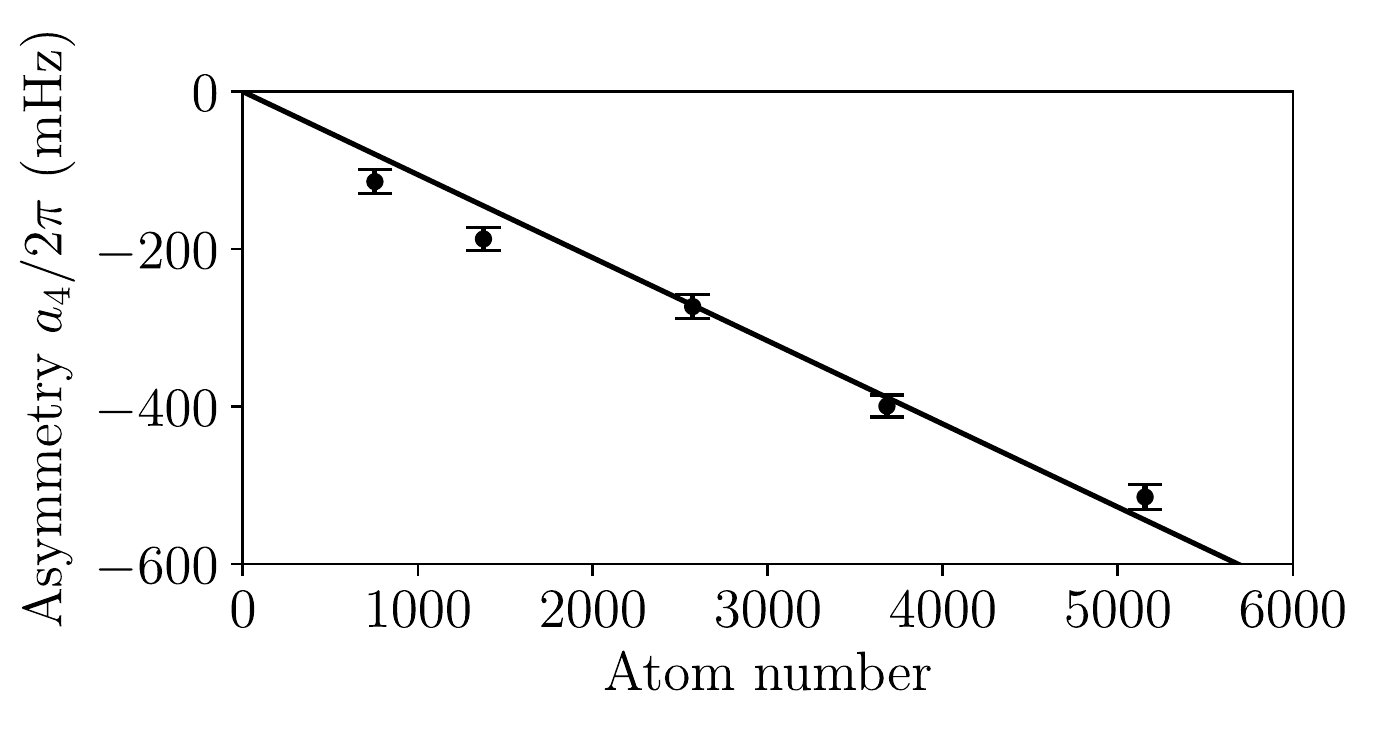}
		\label{densitydependence}
	}
	\hfill
	\vspace*{-6mm}
	\subfloat[]{
		\includegraphics[width=0.40\textwidth]{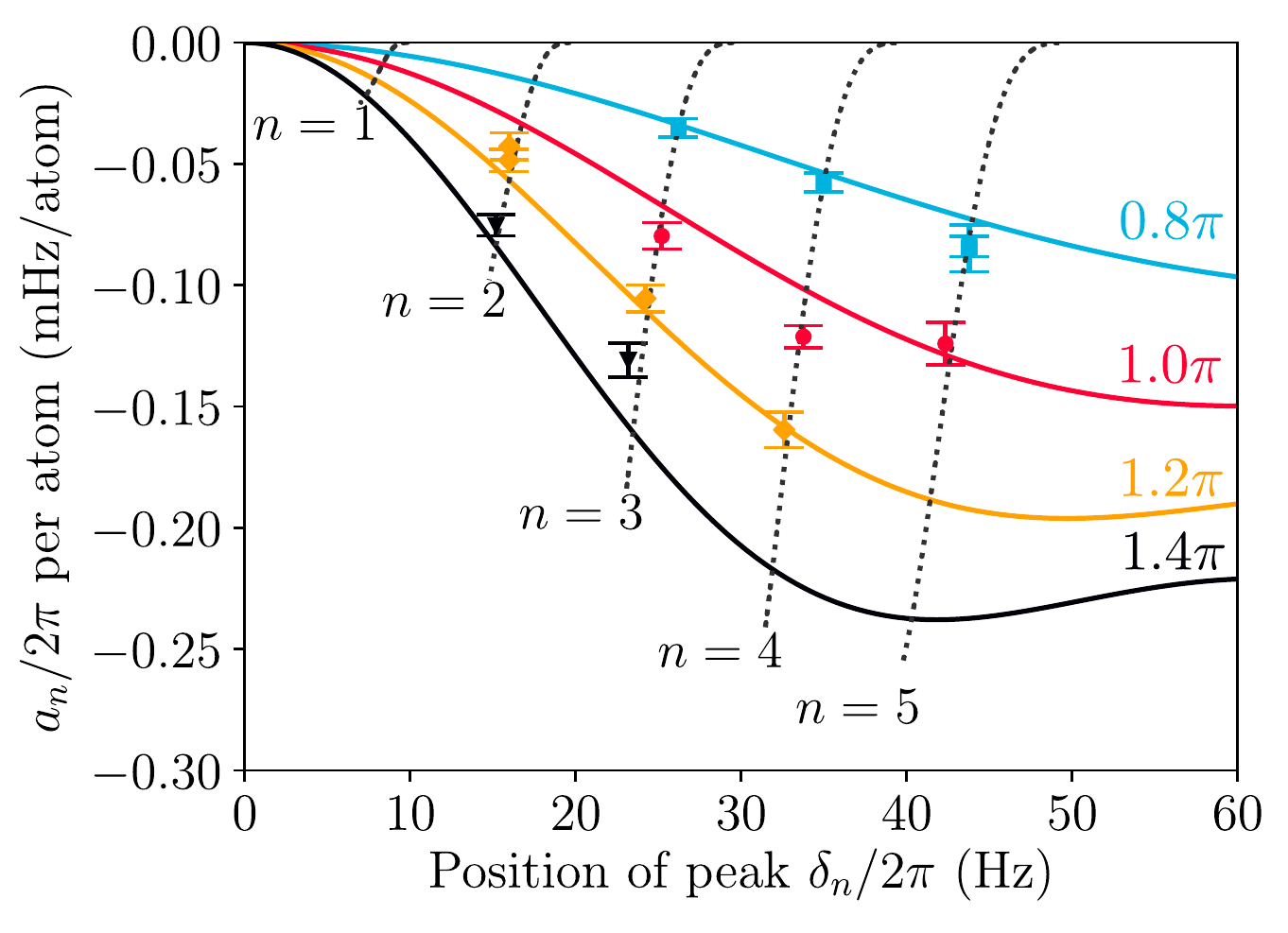}
		\label{asymmetry}}
	\hfill
	\vspace*{-6mm}
	\subfloat[]{
		\includegraphics[width=0.405\textwidth]{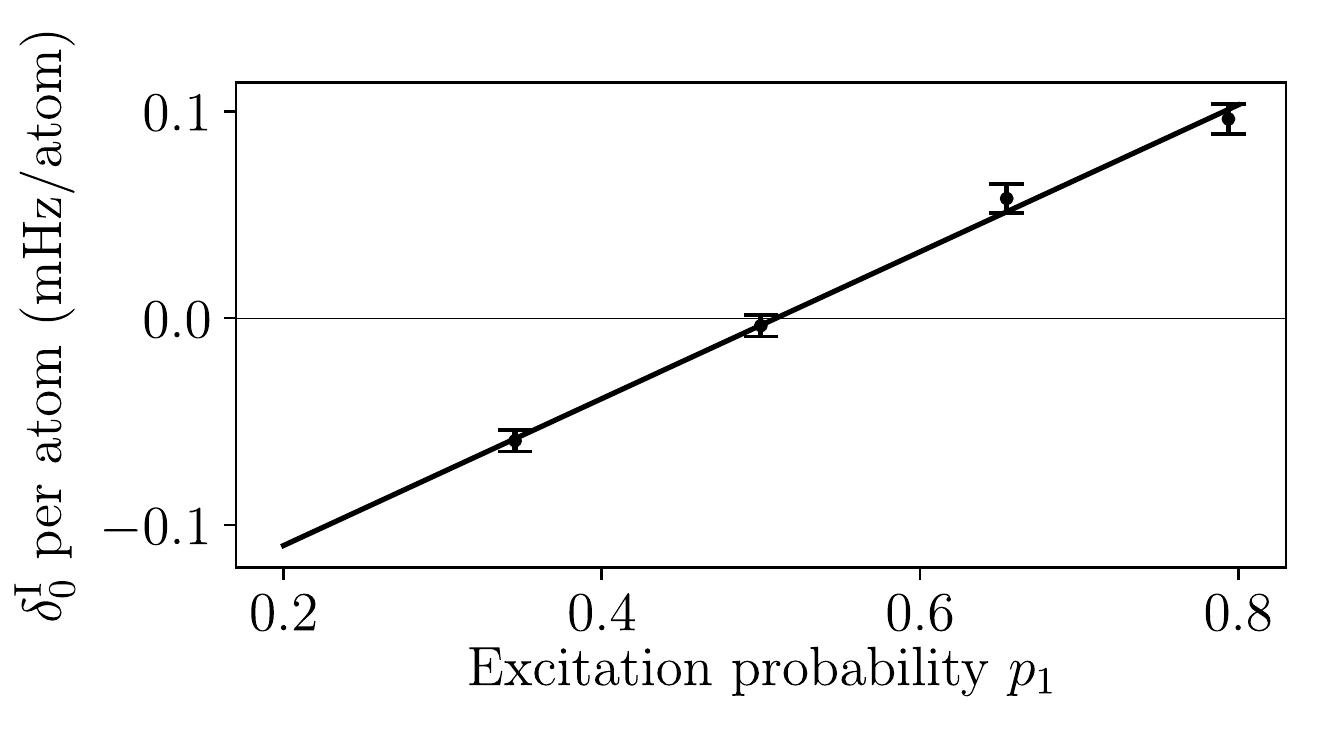}
		\label{excitation}}
		\hfill
			\vspace*{-4mm}
	
	\caption{(a) Asymmetry $a_{4}$ plotted as a function of the atom number with a linear fit. For our Ramsey sequence with $\tau=16$ ms, $T=100$ ms and $\overline{\Omega}\tau=\frac{\pi}{2}$, the $4^\text{th}$ peak is located at $\delta_{4}/2\pi=34$ Hz. (b) Asymmetry $a_{n}$ measured for different $\tau$ with pulse areas $\overline{\Omega}\tau=0.8\pi, 1.0\pi, 1.2\pi, 1.4\pi$ as indicated. All the points are measured with the atom number of $\sim 3000$. The black dashed lines indicate the position of the $n^\text{th}$ peak for each pulse length. (c) Collisional frequency shift of the central peak $\delta_{0}^\text{I}$ is measured for different excitation probability $p_1$ after the first Ramsey pulse. All error bars represent 1-$\sigma$ statistical uncertainties without accounting for instability of experimental parameters.}
	\label{measurements}
\end{figure}

We have defined the interaction parameters through the energy shift due to two-body atomic interactions. While the experimental system of the ${}^{171}\text{Yb}$ clock is not designed to specifically populate two atoms per lattice site, it is possible to describe the overall interaction of a larger number of atoms in the weakly interacting regime as a sum of pairwise interactions \cite{lemke2011}. In the following, we use the effective interaction parameters $\widetilde{V}^{\alpha\beta}$ to represent the total energy shift resulting from interactions with all other atoms in the same site, averaged over the entire lattice. In this way, the model can be tested by comparing the effective $\widetilde{W}$ values obtained through ”$p_1$-based” and ”asymmetry-based” methods. To ensure comparable conditions, following measurements are performed for a constant trap depth of $\sim200E_\text{r}$, and a Ramsey sequence of $T = 100$ ms dark time, enclosed by pulses of $\overline{\Omega}=2\pi\times15.6$ Hz. For the asymmetry-based measurement, the pulse length is chosen as $\tau = 16$ ms for a pulse area of $\frac{\pi}{2}$. We investigate the peaks identified by $n=\pm 4$, which show a significant asymmetry at sufficient signal $P\sim 0.5$ to avoid loss of clock stability due to reduced signal-to-noise ratio.

Measurements of $a_{n}$ are performed by alternately stabilizing the clock laser to the $0^\text{th}$ and $\pm n^\text{th}$ peaks using a three-fold interleaved measurement sequence. First, we investigate the atom number dependence of $a_{n}$ to find the results shown in Figure~\ref{measurements}\subref{densitydependence}. The absence of a significant nonlinearity indicates negligible effects of three-body collisions, which are expected to manifest as a contribution with a quadratic dependence on atom number. The linear dependence allows us to normalize the results to a reference atom number in the following.

The actual measurement of $\widetilde{W}$ through the asymmetry-based method is performed by measuring $a_{n}$ for various $n$ and Ramsey pulse areas $\overline{\Omega}\tau$ and fitting the result for $\widetilde{W}$ using Eq.~(\ref{asymmetryequation}). The measurement results are shown in Figure~\ref{measurements}\subref{asymmetry} together with interpolations according to the fit. The good agreement of the measurement points and theoretical curves confirms the validity of Eq.~(\ref{asymmetryequation}) over a wide range of parameters. The increased magnitude of $a_{n}$ for larger pulse areas $\overline{\Omega }\tau$ reflects that the amplitude of $C_{n}$ has a positive correlation with $\overline{\Omega}\tau$. Dashed lines indicate the position of each peak with varying pulse length $\tau$. Their tilt and curvature are due to the contribution from $\phi(\delta)$. All the points are fitted simultaneously to find $\widetilde{W} = 2\pi \times (0.90 \pm 0.02)$  Hz\footnote{Excess scatter (reduced $\chi^2$ of $5.4$) is attributed to varying trapping conditions, and is not accounted for in the stated statistical uncertainty.} for the typical atom number of $3000$. Since  atoms are  distributed over $500\sim1000$ lattice sites, this represents $3\sim6$ atoms per site.

The alternative measurement of $\widetilde{W}$ is performed by measuring the frequency shift of the central peak $\delta_{0}^\text{I}$ for various $p_1$ and fitting the result using Eq.~(\ref{pwave}) for $\widetilde{W}$. For each measurement, $p_1$ is set to a desired value by changing the pulse areas of both of the Ramsey pulses. $\delta_{0}^\text{I}$ is then determined by measuring the shift of the central peak while alternating between high atom number $N^{(\text{H})}$ and low atom number $N^{(\text{L})}$. The frequency shift $\delta_{0}^\text{I}$ for a specific atom number $N^\text{(T)}$ is extrapolated as $\frac{\delta_{0}^{\text{I}(\text{H})}-\delta_{0}^{\text{I}(\text{L})}}{N^{(\text{H})}-N^{(\text{L})}}N^{(\text{T})}$  where $\delta_{0}^{\text{I}(\text{H})}-\delta_{0}^{\text{I}(\text{L})}$ is the shift measured in the experiment. The measurements results for $\delta_{0}^\text{I}$ are fitted as a linear function of $p_1$ in Figure~\ref{measurements}\subref{excitation}, and the fitting gives $\widetilde{W}= 2\pi\times(1.07\pm0.06)$ Hz for the nominal atom number of $N^{(\text{T})} = 3000$ used in the asymmetry-based measurements.

The stated uncertainties represent only the statistical uncertainty of the contributing measurements and do not account for changes in experimental conditions between measurements. Realistically, we expect about $\sim5$\% variation in the trap depth $D$ for each measurement. This implies $\sim10$\% uncertainty of $\widetilde{W}$, when considering the empirically observed scaling of collisional frequency shifts as $D^2$ or greater. A similar discrepancy will occur if the number of populated lattice sites changes. We thus consider the two methods to show agreement within our measurement precision.

The decomposed description thus succeeds in providing simple explanations for the change of the spectral shape in the presence of atomic interactions. The quantitative relationship between the asymmetry and the interaction parameter $W$ shown in this research reveals that the Ramsey spectra contain information about atomic interactions. Additionally, the asymmetry-based measurement allows extracting information about the strength of the atomic interactions without changing the atom density, which will find useful applications in a case where changing atom number introduces additional effects such as the variation of the populated number of lattice sites.

	\section{Acknowledgments}
This work is supported by JST ERATO Grant Number JPMJER1002 (Japan), by JSPS Grant-in-Aid for Specially Promoted Research Grant Number JP16H06284, and by the Photon Frontier Network Program of the Ministry of Education, Culture, Sports, Science and Technology, Japan.
	\bibliography{mybib}
\end{document}